# Heterostructured hBN-BP-hBN nanodetectors at THz frequencies

*Leonardo Viti[1], Jin Hu[2], Dominique Coquillat,[3] Antonio Politano,[4] Christophe Consejo,[3] Wojciech Knap,[3,5] and Miriam S. Vitiello[1*]*

[1] *NEST, Istituto Nanoscienze – CNR and Scuola Normale Superiore, Piazza San Silvestro 12, Pisa, I-56127*
[2] *Department of Physics and Engineering Physics, Tulane University, New Orleans, LA-70118, USA*
[3] *Laboratoire Charles Coulomb (L2C), UMR 5221 CNRS-Université de Montpellier, Montpellier, F-France*
[4] *Università degli Studi della Calabria, Dipartimento di Fisica, via Ponte Bucci, 87036 Rende (CS), Italy*
[5] *Institute of High Pressure Institute Physics Polish Academy of Sciences Warsaw Poland*

\* miriam.vitiello@sns.it

Keywords: inorganic two-dimensional (2D) materials, van der Walls heterostructures, nano-transistors, photodetectors

**Abstract.** Artificial semiconductor heterostructures played a pivotal role in modern electronic and photonic technologies, providing a highly effective mean for the manipulation and control of carriers, from the visible to the Terahertz (THz) frequency range. Despite the exceptional versatility, they commonly require challenging epitaxial growth procedures due to the need of clean and abrupt interfaces, which proved to be a major obstacle for the realization of room-temperature (RT), high-efficiency devices, like source, detectors or modulators, especially in the far-infrared. Two-dimensional (2D) layered materials, like graphene and phosphorene, recently emerged as a reliable, flexible and versatile alternative for devising efficient RT detectors operating at Terahertz frequencies, with signal to noise ratios SNR ≤ 500. We here combine the benefit of the heterostructure architecture with the exceptional technological potential of 2D layered nanomaterials; by reassembling the thin isolated atomic planes of hexagonal borum nitride (hBN) with a few layer phosphorene (black phosphorus (BP)) we mechanically stacked hBN/BP/hBN heterostructures to devise high-efficiency THz photodetectors operating in the 0.3-0.65 THz range from 4K to 300K with a record SNR = 20000.

Black phosphorus (BP), a recently emerging two-dimensional (2D) van der Waals layered material, is presently stimulating a progressively wider scientific interest for its similarities and, more importantly, differences, with graphene.[1] BP presents a finite, direct band gap ($E_g$) in both its bulk ($E_g$ = 0.35 eV) and monolayer ($E_g \approx 2$ eV) phase, but, in contrast to other common 2D semiconductors ($MoS_2$, $MoSe_2$, $WS_2$, $WSe_2$, etc.), maintains a considerably high mobility (> 1000 $cm^2V^{-1}s^{-1}$ at RT), a clear benefit for nanoelectronic,[2] optoelectronic[3] and photovoltaic[4] applications, requiring a fine control of the device conduction performances. Amongst the most intriguing prospects, the inherently anisotropic BP[5,6] can play a significant role for the realization of switchable logic circuits and radio-frequency field effect transistors (FET),[2] due to its $10^5$ on/off current ratios and large on-state current densities.[7,8] Recently, the strong nonlinear optical response in the visible and mid-infrared ranges has been exploited to devised efficient BP-based fiber-integrated saturable absorbers for ultrafast laser applications.[9] Moreover, the integration of BP crystals in on-chip nanoelectronics has recently led to the implementation of RT photodetectors in the visible, near infrared,[10] telecom[11,12] and THz[13,14] frequency ranges, in this latter case, with performances already suitable for proof-of-principle photonic applications, requiring a few hundred SNR.[14] The realization of RT THz detectors is of utmost importance for the full exploitation of human-centered THz applications, such as medical diagnostics, process control and homeland security. In this context, 2D nano-materials with controllable functionalities offer the intriguing opportunity to design on-chip and scalable detection systems for sensing and imaging applications, providing large quantum efficiency and SNR, and good time stability.

The BP THz detectors demonstrated to date[13,14] exploit the integration of a $SiO_2$-encapsulated single BP flake in an antenna-coupled top-gate (G) field effect transistor (FET). BP provides the unique capability to "engineer" the required detection mechanisms by simply exploiting its structural anisotropy.[13,14] The signal generation occurring within the conductive channel can indeed be designed to arise via different mechanisms (plasma-waves mixing, thermoelectric and bolometric effects), that can be selectively triggered by modifying the arrangement between the FET electrodes and the orientation of the main crystalline directions. It is well known that, because of its strongly corrugated in-plane structure, BP has opposite anisotropies in thermal and electrical conductivities,[5] thus representing an ideal playground for devising different architectures, starting from the same active element.

Despite the conceptual and practical simplicity, the architectures proposed to date have two major limits: (*i*) they don't allow playing much with the device electrical (mobility-constrained) and optical (impedance/capacitance matching) performances; (*ii*) the quite fast degradation of the exfoliated BP flake to oxygen and ambient exposure (moisture, oxygen), prior to encapsulation.[15]

Here, inspired to the artificial semiconductor heterostructure architecture [16] and to the fascinating capabilities of Van der Waals (vdW) heterostructures [17-20] we embed a BP flake within a natural semiconductor heterostructure formed by multilayered hexagonal boron nitride (hBN) crystals, to devise hBN/BP/hBN heterostructured THz photodetectors having high optical response, and an extremely good time-dependent electrical stability.

hBN, a III-V compound with an energy band gap of $\approx$ 5.2-5.4 eV[21] constitutes an ideal encapsulating material for BP. Being impermeable to gases and liquids it provides permanent shielding to ambient exposure, leading to extremely air-stable devices.[22] Furthermore, it enables reaching record mobility (up to $\approx$ 1350 cm$^2$V$^{-1}$s$^{-1}$ at room temperature[23] and $\approx$ 6000 cm$^2$V$^{-1}$s$^{-1}$ for T < 30 K[20]) thanks to its flatness and its compatibility with honeycomb structure. In contrast with oxide-encapsulated structures,[8] when stacked over other layered crystals, hBN indeed forms clean and inert interfaces, preventing charge traps or dangling bonds.[24] Under this configuration, hBN can also be seen as a valuable *ready-to-use* gate dielectric, due to its high breakdown voltage ($\approx$ 1 V nm$^{-1}$, depending on the number of layers and on the carrier type[25]) and the sufficiently large dielectric constant (3.5-4),[26,27] which in turn allows high gate-to-channel capacitance value required for tuning the device transport and optical properties. Moreover, it is in principle capable to provide a large mobility increase at low temperatures (≈ an order of magnitude)[16,20] with respect to any oxide-surrounded device (≈ a factor of 2).[6,28]

The hBN/BP/hBN heterostructured nanodevices were devised over a high resistivity silicon substrate covered with a 300 nm $SiO_2$ cap-layer. A first set of hBN flakes were mechanically exfoliated onto the clean $SiO_2$ surface and their thickness measured with an atomic force microscope (AFM). 40 nm thick hBN flakes having lateral dimensions of 20x20 μm$^2$ were selected to facilitate subsequent alignment steps. BP flakes were then transferred onto the lower hBN layer with a deterministic dry transfer process, the polyvinylalcohol (PVA) method,[29] which ensures an extremely clean interface between the two materials.

Figures 1(a-b) show the scanning electron micrograph (SEM) and AFM images of one of the fabricated stacks, respectively. The source (S) and drain (D) contacts were defined through a combination of electron beam lithography (EBL) and metal deposition on the BP flakes before the deposition of the hBN top layer; this approach should in principle reduce the contact resistance with respect to one-dimensional electrical contacts obtained on the 2D materials via diagonal etching of the heterostructure (See Methods).[18,30] To ensure an optimal radiation feeding,[13] we patterned the S electrode in shape of the arm of a planar bow-tie antenna and the D electrode in the shape of a straight line (**Figure 1c**). The S/D metal deposition was performed in two separated lithographic steps (See Methods). This double-step evaporation procedure (**Figure 1d**) allows realizing thick antennas and mechanically robust bonding pads, while keeping low (< 1 fF) the parasitic capacitances between the FET electrodes. The top hBN flake, serving as passivation layer and simultaneously as top-gate isolating dielectric, was then transferred on the hBN-BP stack using the deterministic PVA method. The gate (G) electrode, patterned in shape of a split bow-tie arm (**Figure 1c**),[31] was then deposited on the top of the hBN-BP-hBN heterostack.

We devised two samples exploiting a different crystallographic orientation of the BP flake with respect to the FET channel, while keeping unchanged the antenna architecture (see Methods). Before the fabrication of S and D contacts, the in-plane orientation of the transferred BP flakes was determined by means of Raman spectroscopy experiments, performed following the method described in Refs. [6, 14] (see Supporting Information). Sample A has been electrically contacted along the 45° direction in-between the armchair (x) and the zigzag (y) axis, whereas sample B has been contacted along the zigzag (y) direction. Such a difference is expected to affect the transport and optical response of the two samples.[14] The AFM topographic image of the processed devices (**Figure 1d**) revealed the presence of a narrow (≈ 30 nm) vertical gap between the top hBN layer and the BP channel, induced by the inherent thickness of S and D electrodes. This gap affects the gate to channel capacitance ($C_{gc}$) that, in the present case, has been simulated to be $C_{gc} \approx 200$ aF (3D FEM, COMSOL Multiphysics), resulting in a five times smaller capacitance per unit area (~ 160 µF/m$^2$) with respect to previously reported SiO$_2$ encapsulated BP FETs.[13,14]

Figures 2a-b shows the gate bias dependence of the channel resistance trend in sample A and B, respectively, which allows inferring the transistor transconductance $g_m$. In order to prevent leakage currents or damages to the isolating upper hBN layer, $V_G$ was kept between -15 V and + 15 V.

The detectors were optically tested at RT by using two different tunable THz electronic sources, operating, respectively, in the frequency ranges 265-375 GHz and 580-640 GHz (see Methods). Room-temperature photoresponse experiments were performed in both photovoltaic (PV) and photoconductive (PC) modes. In the first approach, the photovoltage Δu was measured at the D electrode of the FET using a voltage preamplifier in series with a lock-in, while keeping S grounded and the G terminal connected to a voltage generator. In the second approach, which is a purely *dc* measurement, a direct S-D current ($I_{SD}$) arises in the BP channel after applying a fixed S-D dc voltage ($V_{SD}$). $I_{SD}$ fluctuations can then be measured at the drain side as a function of the optical power and of the impinging frequency.

The photovoltage values extrapolated via the PV experiments are shown in Fig.2c and Fig.2d for samples A and B, respectively (see Methods). The comparison among the $\Delta u$ values measured when the beam was impinging on the detector and when it was blanked ($\Delta u_b$, inset of Fig.2c), allows evaluating the signal-to-noise ratio that, in the case of sample A at 294 GHz, is SNR ≈ 3200, i.e. significantly larger that achieved in the case of BP photodetectors, confirming the superior performances ensured by the hBN/BP/hBN interface. The extrapolated responsivity for sample A remains almost constant as a function of $V_G$ reaching $R_v$ = 1.7 V/W at 294 GHz and $R_v$ = 0.2 V/W at 593 GHz (**Figure 2c**). The same experiments on sample B, allow achieving an identical noise level ($\Delta u_b$ ≈ 10 nV), with a slightly smaller responsivity $R_v$ = 0.3 V/W at 295 GHz which increases up to 0.9 V/W at 627 GHz (**Figure 2d**). The almost flat photoresponse (**Figures 2c-d**) is well compatible with the 1.5 $I_{on}/I_{off}$ ratio visible in the transport characteristics (**Figures 2a-b**) which reflects the expected optical behaviors. Surprisingly, the two samples show different frequency-dependent photoresponse: while indeed the responsivity of sample A is decreasing at larger frequency, sample B has a significantly larger sensitivity at 627 GHz. This is ascribed to the superposition of different physical mechanisms, simultaneously activated within the channel. Indeed, thermoelectric, plasma-wave or bolometric effects can contribute to the overall photoresponse and their interplay is strictly dependent from the device architecture. To infer the physical mechanisms at the origin of the detection process one can compare the photovoltage and photocurrent measurements, which provide an unambiguous way to identify the related physical dynamics.[14]

The thermoelectric effect arises when a temperature gradient is generated within a non-zero Seebeck coefficient material. This is expected to occur in the present case since the devised asymmetric antenna couples the THz radiation more efficiently to the S electrode than to the D electrode, leading to different thermal distributions at the two ends of the BP channel. As a consequence of the carriers (holes) diffusion from the hot to the cold FET side, a net current flows toward the D electrode. Thus, if a PC measurement is performed, the thermoelectric current ($I_T$) is negative regardless the sign of the applied $V_{SD}$.

Plasma-wave effects are usually triggered by a specific degree of asymmetry provided by the antenna geometry, the channel material, or by a longitudinal electric field along the S-D channel. In a PC measurement, the dominant asymmetry is induced by the applied $V_{SD}$, the plasma-wave-related current ($I_{pw}$) will flow in the same direction of $V_{SD}$, keeping its sign. Conversely, the bolometric effect is caused by a channel conductivity change induced via heating or cooling processes. The amplitude of the bolometric current ($I_B$) is proportional to the bolometric coefficient $\gamma = d\sigma/dT$. In the case of positive $\gamma$, the bolometric contribution to the current sums up with the same sign to any electric field driven source-drain current: if $V_{SD}$ is positive, $I_B$ follows its sign.

Summarizing the former considerations in a schematic formula, the *dc* current flowing through the detectors when measured in PC mode can be written as:

$$\begin{cases} I_{SD\text{-}on}^{(+)} = I_{SD\text{-}off}^{(+)} + I_{pw} + I_B - I_T; & (V_{SD} > 0) \\ I_{SD\text{-}on}^{(-)} = I_{SD\text{-}off}^{(-)} - I_{pw} - I_B - I_T; & (V_{SD} < 0) \end{cases} \quad (1)$$

Where the (+) and (-) superscripts indicate the sign of $V_{SD}$. Equation (1) can be also written as:

$$\begin{cases} \frac{I_T}{\sigma} = -\frac{\Delta u^{*(+)} + \Delta u^{*(-)}}{2} & (2) \\ \frac{I_{pw} + I_B}{\sigma} = \frac{\Delta u^{*(+)} - \Delta u^{*(-)}}{2} & (3) \end{cases}$$

where the photoconductive voltage $\Delta u^* = (I_{SD-on} - I_{SD-off})/\sigma^{-1}$ accounts for the difference between the photocurrent detected when the THz beam is impinging on the heterostructure ($I_{SD-on}$) and when the beam is blanked ($I_{SD-off}$) and where σ is the *dc* conductivity. Equation (2) is the thermoelectric photovoltage while equation (3) is the sum of the plasma-wave and bolometric voltage. Therefore, photocurrent measurements at positive and negative $V_{SD}$ allow separating the thermoelectric contributions from alternative physical dynamics.

Figures 2e and 2f show the direct comparison between the responsivity values extracted via the PV methods and the photovoltage contributions of equations 2 (thermoelectric $\Delta u_{th}^*$) and 3 (plasma-wave/bolometric $\Delta u_{pwB}^*$) for samples A and B, respectively. From the above comparison we can extract the correlation coefficient (*Cc*), which can be assumed as an indicator of the degree of similarity (linear dependence) of two variables: a 100% *Cc*(*f, g*) means that the relationship between two functions *f* and *g* is perfectly described by a linear equation (see Methods). The analysis of Figs 2g allows extracting $Cc(\Delta u_{th}^*, R_v) = 60\%$ and $Cc(\Delta u_{pwB}^*, R_v) = 50\%$, indicating that, in sample A, the effects are simultaneously contributing to the overall signal. Conversely, in the case of sample B (Fig. 2h) we estimate $Cc(\Delta u_{th}^*, R_v) = 70\ \%$ and $Cc(\Delta u_{pwB}^*, R_v) = 0.5\ \%$, thus indicating that THz detection is fully mediated by the thermoelectric effect.
This conclusion is in general agreement with our previously experimental observations on 45° oriented-BP based photodetectors,[14] demonstrating that the BP flake is mainly governing the overall electrical/thermal management.

In order to test the low-temperature performance of the devised detectors, we employed a helium refrigerated cryostat, and varied the heat sink temperature (T) from 4 K to 300 K. The optical setup was similar to the one used in the RT experiments, with the difference that the collimated (not focused) THz beam, after being partially attenuated by the cryostat window, was sent to the samples with a reduced incoming optical intensity of ~ 3.1.

**Figure 3a-b** shows the temperature dependence of the $R_v$ plot, collected at 295 GHz. A progressively larger increase of the photoresponse while decreasing T is achieved. Moreover, a visible gate bias dependence in the optical response appears at temperatures below 50K. This can be ascribed to the low-temperature enhanced capability of the G electrode to modulate the charge density when the number of available carriers for conduction decreases. In sample A, $R_v$ reaches a maximum value of 38 V/W at 4 K, when the G electrode is biased at +4 V. For sample B, $R_v$ shows two maxima ≈ 20 V/W corresponding to $V_G$ = -3 V and $V_G$ = -15 V.

Under the assumption that the noise figure is dominated by the thermal noise, we determined the detectors noise equivalent power (NEP) as the ratio between the Johnson-Nyquist noise spectral density $N_{th} = (4K_B\sigma^{-1}T)^{1/2}$, where $K_B$ is the Boltzmann constant, and the responsivity. The results reported in Figure 3 c-d show a minimum NEP~ 100 pW/(Hz)$^{1/2}$ at 4 K, for both devices. Remarkably, the correspondent signal-to-noise ratio reaches 20000, a record value for any van der Waals 2d-material based THz detectors. [32,33]

In order to have a deeper insight on the low temperature dynamics, we performed photovoltage experiments in PV mode while varying the output frequency *f*, in the range 265-375 GHz and T in the range 4-298 K and while keeping fixed $V_G$ = 0V. Interestingly, we discover that the spectral response of samples A and B shows a very different temperature dependence (**Figure 4 a-b**).

In sample B, by varying the temperatures, the signal amplitude increases but the peak frequency position remains unchanged (**Figure 4b**). Conversely, sample A shows a visible temperature dependence of the main peak frequency position. This effect is made evident in Figure 4a, which clearly shows the red-shift occurring when the temperature varies in the 100 K - 300 K range.

In sample A, the top gate field-effect hole mobility[13] $\mu = g_m L_G^2/C_{gc}V_{SD}$, is almost constant, (~ 550 cm$^2$V$^{-1}$s$^{-1}$), in the temperature range 4 K - 100 K, then decreases to 210 cm$^2$V$^{-1}$s$^{-1}$ at room temperature (**Figure 4c**). In sample B, μ follows a rather similar trend, but reaches higher values (up to ≈ 1050 cm$^2$V$^{-1}$s$^{-1}$) at low temperature (**Figure 4d**), in qualitative agreement with previous reports on p-type BP-based field effect transistors.[18,20,24]

The comparison between the spectral response and the mobility trend of sample A, unveils that the red-shift in the spectral response occurs in the same temperature range where its mobility starts to decrease. On the other hand, though sample B presents an even larger mobility variation, its frequency dependence is practically unaffected by temperature changes. This difference can be attributed to the dissimilar mechanism of signal generation within the two detectors. Nevertheless, this conclusion leaves open a question about the low temperature responsivity increase. Indeed, the thermoelectric contribution to Δu is expected to follow the trend of the Seebeck coefficient ($S_B$), which is expected to decrease at low temperature.[34]

In order to unveil the physical origin of THz detection we analyzed the amplitude profile of a resonance frequency peak (dashed lines in **Figures 4 a-b**, corresponding to ≈ 310 GHz for sample A and to ≈ 311 GHz for sample B) as a function of temperature. **Figures 4 e-f** show the measured $R_v(T)$ of sample A and B, in the range 4 K - 300 K (left vertical axis), in comparison with the expected value of the bolometric contribution to the photovoltage ($V_B$ – right vertical axis). The latter has been estimated from the ratio between the bolometric coefficient γ (defined earlier in the text) and the channel conductivity σ:[14]

$$V_B = \frac{\gamma}{\sigma} = \frac{1}{\sigma} \cdot \frac{d\sigma}{dT} \qquad (4)$$

In both cases, we found $V_B > 0$ (the conductivity increases with temperature) with a monotonic decrease at increasing T. Below 50 K, $R_v(T)$ matches the $V_B(T)$ trend, meaning that the low temperature detection can be reasonably ascribed to the onset of bolometric effect, which becomes the dominant contribution.

Interestingly, the $R_v(T)$ plot shows a broader peak around 150 K (particularly evident in the case of sample A) which is in agreement with the expected trend of the Seebeck coefficient for a few-layer BP devices. The responsivity analysis as a function of gate voltage, frequency and operating temperature allowed us to identify a device (sample A) in which the superposition and interplay of the three main physical mechanisms for detection offer an intriguing benchmark for the understanding and future development of BP-based cooled and room temperature THz detectors based on 2D nanomaterials and heterostructures.

**Methods**

*Fabrication.* The metal deposition of the source and drain electrical contacts was performed in two separated steps: the portion of the contacts in close proximity with the BP flake was covered with 5/50 nm Ni/Au, while the more external portions (at ~ 5 μm distance) with a 10/150nm Cr/Au layers sequence (see Fig.1d). This double-step evaporation procedure allows realizing thick antennas and mechanically robust bonding pads, while keeping low (< 1 fF) the parasitic capacitances between the FET electrodes. A second hBN flake, serving as passivation layer and, simultaneously, as top-gate isolating dielectric, was then transferred on the hBN-BP stack using the deterministic PVA method. The gate (G) electrode, patterned in shape of a split bow-tie arm[31] (Fig.1c), was deposited on the top of the hBN-BP-hBN stack via thermal evaporation of a 10/150 nm Cr/Au layer. We devised two set of samples exploiting a different crystallographic orientation of the BP flake. In both cases, identical split bow-tie antennas having a total length of 500 μm have been exploited. All the relevant geometrical dimensions were kept equal to allow for an immediate comparison between the devices electrical and optical performances: the channel length is $L = 2$ μm, the gate length is $L_G = 850$ nm, the BP thickness is 25±2 nm (corresponding to about 40±4 layers) and the top hBN layer is 50-55 nm thick. The main difference between samples A and B is the flake orientation with respect to the active channel S-D axis.

*Optical characterization.* The beam, mechanically chopped at 619 Hz, was focused by a set of off-axis f/# = 1 parabolic mirrors, reaching a focal beam spot diameter of ≈ 4 mm. The optical power impinging on the samples, calibrated as a function of frequency by a power meter, varies between 260 μW and 400 μW. The extrapolated photovoltage value $\Delta u = 2.2*LIA/G$,[13] where LIA is the lock-in signal and G the pre-amplifier gain. The detector responsivity ($R_v$), defined as the ratio between the detector photoresponse and the input power, is determined by the relation $R_v = (\Delta u/P)*(S_t/S_a)$, where P is the total power at the focal point, $S_t$ is the beam spot area and $S_a$ is the active area of the device. Since in the presented experiments $S_a$ is smaller than the diffraction limited area ($S_\lambda = \lambda^2/4$), $S_a$ has been assumed equal to $S_\lambda$.

*Correlation analysis of the main physical mechanisms.* The correlation between two measured variables has been computed as the Pearson correlation coefficient (defined as *Cc*), which is an estimation of their linear dependence; given two variables *f* and *g*, each with N observations, *Cc* is calculated as the mean of the product of the standard scores for *f* and *g*:

$$Cc(f,g) = \frac{1}{N-1}\sum_{i=1}^{N}\left(\frac{f_i - \overline{M_f}}{D_f}\right)\left(\frac{g_i - \overline{M_g}}{D_g}\right)$$

Where $M_f$ and $M_g$ are the mean values and $D_f$ and $D_g$ are the standard deviations for *f* and *g*, respectively. The extrapolated *Cc* is a number in the range (-1;+1). A value of 1 implies that a linear equation describes the relationship between *f* and *g* perfectly, with all data points lying on a line for which *f* increases as *g* increases. A value of −1 implies that all data points lie on a line for which *g* decreases as *f* increases. A value of 0 implies that there is no linear correlation between the variables.

**Acknowledgments**

The authors acknowledge support from the European Union Seventh Framework Programme grant agreement n° 604391 Graphene Flagship, the European Union through the MPNS COST Action "MP1204 TERA-MIR Radiation: Materials, Generation, Detection and Applications", the ANR P2N NADIA "Integrated NAno-Detectors for terahertz Applications" (ANR-13-NANO-0008), the National Science Poland Centre (DEC-2013/10/M/ST3/00705), the NSF/LA-SiGMA program under award #EPS-1003897. JH thank Z. Mao for helpful discussions.

**Figure Captions**

**Fig.1: Device Fabrication.** (**a**) Scanning electron micrograph (SEM) of few-layers BP flakes transferred on top of exfoliated hBN crystals. (**b**) Atomic force microscope (AFM) image of the area enclosed by the yellow square in (**a**). The height profile of the stack is reported, showing a BP thickness of ~ 25 nm, corresponding to 40 monolayers. (**c**) Optical microscope image of one of the fabricated device. The bow arm length is L = 250 μm. (**d**) AFM tomographic image of the top-gate field effect transistor.

**Fig.2: Room temperature characterization.** (**a,b**) Source-drain resistance ($R_{SD}$) as a function of gate voltage ($V_G$) for sample A (a) and B (b), respectively. (**c,d**) Photovoltage (Δu) measured at different operating frequencies, plotted as a function of $V_G$ for sample A(c) and B(d); inset: noise level obtained when the 295 GHz beam is blanked. (**g,h**) Sample A,B: left vertical axis, $R_v$ as a function of frequency as obtained from photovoltage (PV) measurements. Right vertical axis: thermoelectric (purple) and plasma-wave/bolometric (green) photovoltage contributions measured via photocurrent (PC) experiments, as estimated from equations (2) and (3).

**Fig.3: Detection performance at low temperature.** (**a,b**) Sample A,B: $R_v$ as a function of $V_G$ measured at different temperatures with a 295 GHz incoming frequency. (**c,d**) Sample

A,B: noise equivalent power (NEP) extracted as the ratio between the thermal Johnson-Nyquist nose spectral density ($N_{th}$) and the responsivity.

**Fig.4: Temperature dependence analysis.** (**a,b**) Sample A,B: normalized $R_v$ chart as a function of incoming frequency (left vertical axis) and temperature (horizontal axis), measured with $V_G = 0$ V. (**c,d**) Sample A,B: field effect hole mobility as a function of temperature. (**e,f**) Sample A,B: temperature dependence of $R_v$ (left vertical axis) measured following the signal peaks at ≈ 310 GHz (green dashed line in panel **a,b**), in comparison with the theoretical bolometric contribution ($V_B$) to the photovoltage obtained from equation (4), right vertical axis. The expected trend of the thermoelectric contribution, proportional to the Seebeck coefficient ($S_B$), is shown in grey. Lines are guides to the eye.

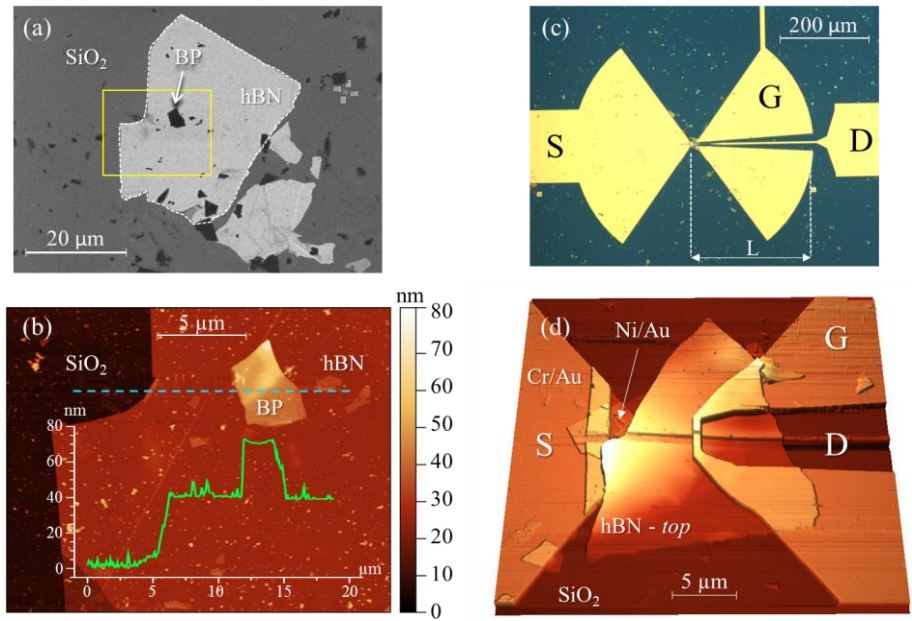

**Figure 2**

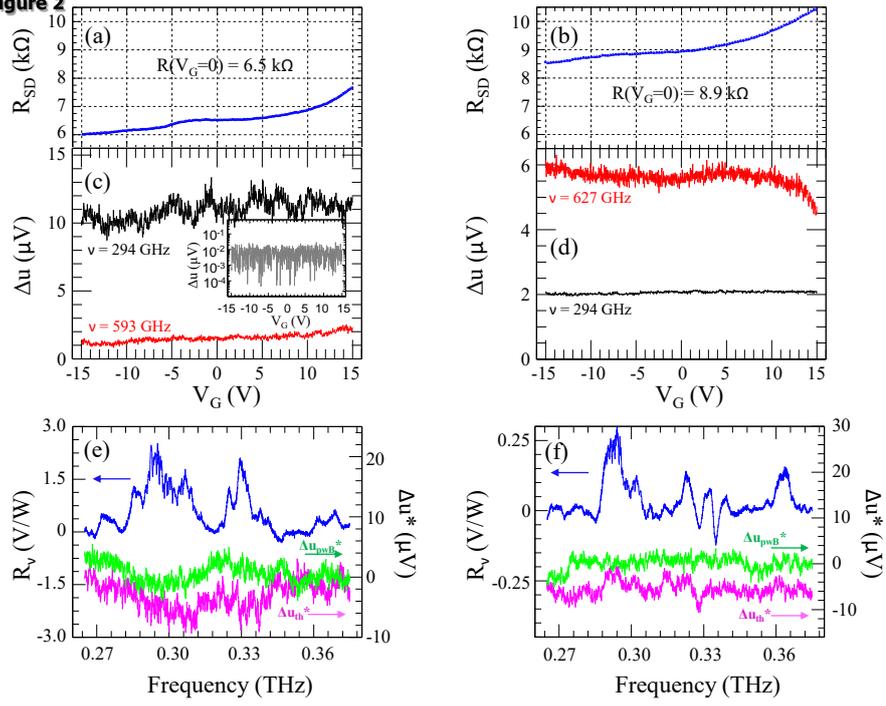

**Figure 3**

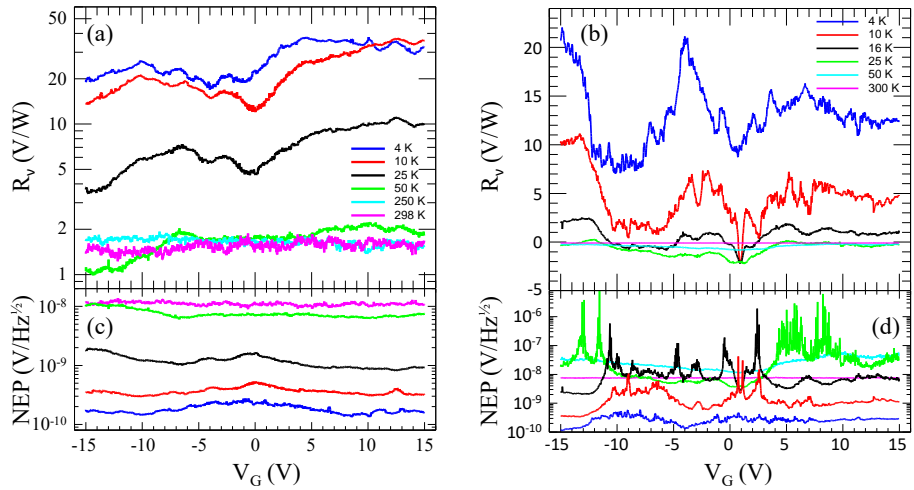

**Figure 4**

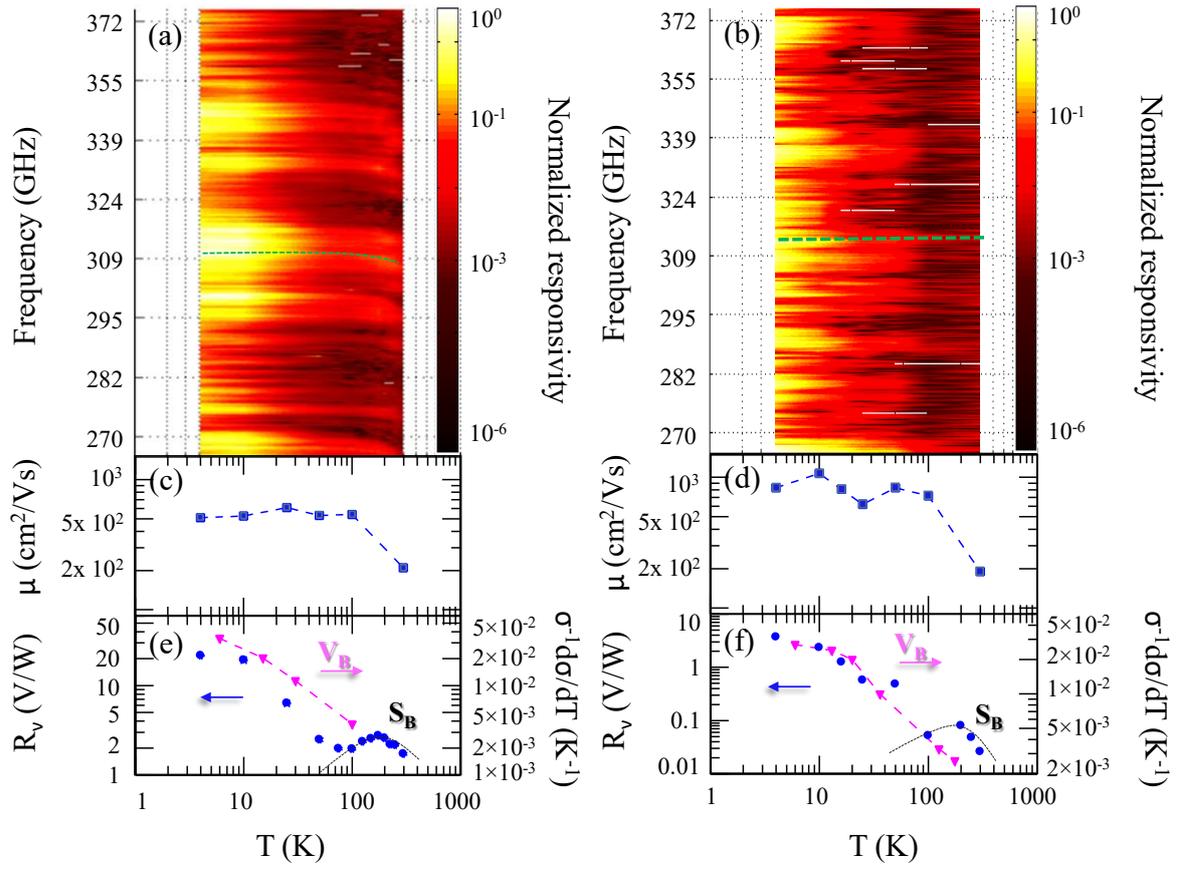